\documentclass{cccg16}

\usepackage{amssymb, amsmath}
\usepackage{graphicx}
\usepackage[hidelinks]{hyperref}
\usepackage[utf8]{inputenc}
\usepackage[T1]{fontenc}
\usepackage{lmodern}
\usepackage{hhline}
\usepackage{wrapfig}
\usepackage{subfig}
\usepackage{listings}
\usepackage{courier}
\usepackage{float}
\usepackage{color}
\usepackage{dcolumn}

\usepackage{lipsum}

\DeclareMathOperator{\sign}{sign}

\title{On the Precision to Sort Line-Quadric Intersections}
\author{Michael Deakin \and Jack Snoeyink\thanks{School of Computer
    Science, University of North Carolina at Chapel Hill, {\tt
      mfdeakin@cs.unc.edu}, {\tt snoeyink@cs.unc.edu}}}

\index{Deakin, Michael}
\index{Snoeyink, Jack}

\newcolumntype{d}[1]{D{.}{\cdot}{#1} }

\begin{document}
\thispagestyle{empty}
\maketitle

\begin{abstract}
  To support exactly tracking a neutron moving along a given line
  segment through a CAD model with quadric surfaces, this paper
  considers the arithmetic precision required to compute the order of
  intersection points of two quadrics along the line segment. When the
  orders of all but one pair of intersections are known, we show that a
  resultant can resolve the order of the remaining pair using only
  half the precision that may be required to eliminate radicals by
  repeated squaring. We compare the time and accuracy of our technique
  with converting to extended precision to calculate roots.
\end{abstract}

\section{Introduction}
In this work, we are concerned with ordering the points of
line-quadric intersections in 3 dimensions, where the inputs are
representable exactly using $w$-bit fixed-point numbers.  We will
actually use floating point in storage and computation, but our
guarantees will be for well-scaled inputs, which are easiest described
as fixed-point.  A {\it representable point}~$q$ or {\it representable
  vector}~$v$ is a $3$-tuple of representable numbers $(x, y, z)$. The
line segment from point~$q$ to~$q+v$ is defined parametrically for
$t\in [0,1]$ as $\ell(t)=q+tv$; note that there may be no
representable points on line $\ell$ except its endpoints (and even
$q+v$ may not be representable, if the addition carries to
$w+1$~bits.)

A quadratic is an implicit surface defined by its 10 representable coefficients,
\begin{align*}Q(x, y, z)=q_{xx} x^2 &+ q_{xy} xy + q_{xz} xz + q_x x + \dots \\
&+ q_{zz} z^2 + q_{z} z + q_c = 0.
\end{align*}
For more accuracy, we can allow more precision for the linear and
quadratic coefficients, since we will need $3w$~bits to exactly
multiply out the quadratic terms, or we can use a representable
symmetric $3{\times} 3$ matrix~$M$, a representable vector~$v$, and a
$3w$-bit constant~$R$ to give a different set of quadrics $\tilde Q(p)
= (p-v)^TM(p-v) = R$ that is closed under representable translations
of~$v$. Whichever definition of quadrics is chosen, the parameter
values for line-quadric intersections are the roots of $Q(\ell(t))=0$,
which can be expressed as a quadratic $at^2+2bt+c=0$ whose
coefficients can have at most $3w+4$~bits.  (Four carry bits suffice
to sum the $3w$-bit products; $w=16$ allows exact coefficient
computation as IEEE 754 doubles; $w=33$ as pairs of doubles.)

These definitions are motivated by a problem from David Griesheimer,
of Bettis Labs: rather than tracking a particle through quadric
surfaces in a CAD model, would it be more robust to compute the
intervals of intersections with a segment?  We compare three methods
to order line-quadric intersections.  Our methods, particularly the
third, are developed and tested for the case where only one pair of
roots has a difference that is potentially overwhelmed by the rounding
errors in the computation. We comment at the end how to handle pairs
of quadric surfaces that have more than one pair of ambiguous roots.

\section{Methods}
This section outlines three methods---Approximate Comparison, Repeated
Squaring, and Resultant---to sort the intersections with two quadrics,
$Q_1$ and $Q_2$, with a given line $\ell(t)$, or equivalently, the
roots of two quadratics, $a_1t^2+2b_1t+c_1=0$ and
$a_2t^2+2b_2t+c_2=0$.  For each, we evaluate correctness, precision,
and floating-point arithmetic operations (FLOPs) required.

\subsection{Approximate Comparison}
The approximate comparison method computes, for $i\in\{1, 2\}$, the
roots~$r_i^\pm=({b_i\pm\sqrt{b_i^2-a_ic_i}})/{a_i}$ approximately by
computing each operation in IEEE 754 double precision or in extended
precision.  Actually, to avoid subtractive cancellation, we calculate
one of the two roots as $r_i^{-\sign
  b_i}=-c_i/({b_i+(\sign{b_i})\sqrt{b_i^2-a_ic_i}})$.  The order of
any two chosen approximate roots can be calculated exactly
as~$\sign(r_1^\pm-r_2^\pm)$.

The rounding of floating point arithmetic means that even with
representable input, the correct order is not guaranteed unless we
establish a gap or separation theorem (which are also established
using resultants~\cite{brownawell2009lower,emiris2010dmm}) and compute
with sufficient precision. Determining this precision is a
longstanding open problem~\cite{demaine33open}.  Without a guarantee,
this method requires very little computation.  Computing both roots
takes $12$ FLOPs, with one more to compute the sign of the difference.
Moreover, the roots can be reused in a scene of many quadrics.

We also use extended precision, where the multiplications and addition
in the discriminants are calculated with $6w$ bits, square root and
addition at $12w$ bits, and divisions at $24w$ bits.  To actually
perform the comparison, one final subtraction is required at 24 times
the initial precision -- 1 FLOP, with an initialization cost of 10
FLOPs per quadric intersecting the line.

\subsection{Repeated Squaring}
The repeated squaring method computes $\sign(r_1^\pm-r_2^\pm)$ by
algebraic manipulations to eliminate division and square root
operations, leaving multiplications and additions whose precision
requirements can be bounded.  It uses, for $x\ne 0$, the property that
$\sign(y)=\sign(x)\sign(x\cdot y)$.  Divisions can be removed
directly, since $\sign(r_1^\pm-r_2^\pm)=\sign(a_1 a_2)\sign(a_1 a_2
(r_1^\pm-r_2^\pm))$.  One square root can be eliminated by multiplying
by $r_1^\pm-r_2^\mp$, giving~$\sign(a_1 a_2)\sign(a_1 a_2
(r_1^\pm-r_2^\mp))\cdot\sign(a_1^2 a_2^2 (r_1^\pm - r_2^\pm) (r_1^\pm
- r_2^\mp))$.  When simplified, the final sign is computed
from~$a_2^2b_1^2-2a_1a_2^2c_1+2a_1^2a_2c_2-a_1a_2b_1b_2\pm
\sqrt{(a_1a_2b_2-a_2^2b_1)^2(b_1^2-4a_1c_1)}$.

The expression under the radical is correctly computed with $8\times$
the input precision; the remaining expression can be evaluated to a
little more than $4\times$ input precision in floating point, or can
be evaluated in fixed point in $8\times$ input precision by isolating
the radical and squaring one last time.

This method not only requires high precision, but also a large number
of FLOPs.  Computing the unambiguous sign of the difference of the
roots requires 15 FLOPs total, and correctly computing the final sign
requires another 24 FLOPs.  Unfortunately, many of the computed terms
require coefficients from both polynomials; only the discriminants,
squares, and products can precomputed, which reduces the number of
FLOPs by 14.  This brings us to 25 FLOPs per comparison, with an
initialization cost of 14 FLOPs per quadric.

Note that this method uses our assumption that we know
$\sign(r_1^\pm-r_2^\mp)$ when computing $\sign(r_1^\pm-r_2^\pm)$, but
we can learn this from a lower precision test against $-b_2/a_2$,
since $r_2^- \le -b_2/a_2 \le r_2^+$.

\subsection{Resultant}
This method was previously described in \cite{fastaccuratefp}, but a
description is included here for completeness.

The resultant method computes the order of two intersections from the
resultant for their polynomials, which can be written as the
determinant of their Sylvester Matrix~\cite[Section~3.5]{cheeyap}.
The general Sylvester Matrix for polynomials~$P(t)=p_m t^m + \dots +
p_0$ and~$Q(t)=q_n t^n + \dots + q_0$ is defined as in Equation
\ref{eq:sylv}.

\begin{equation}
  res(P, Q)=\begin{pmatrix}
    p_m & \dots & & p_0 & 0 & & 0\\
    0 & p_m & \dots & & p_0 & & 0\\
    & \ddots & & & & \ddots\\
    0 & 0 & & p_m & \dots & & p_0\\
    q_n & \dots & & q_0 & 0 & & 0\\
    0 & q_n & \dots & & q_0 & & 0\\
    & \ddots & & & & \ddots\\
    0 & & q_n & \dots & & q_0\\
  \end{pmatrix}
  \label{eq:sylv}
\end{equation}

The resultant is also the product of the differences of $P$'s roots,
$a_1$, \dots, $a_n$, and $Q$'s roots, $b_1$, \dots, $b_m$, as in
Equation~\ref{eq:resultant}.~\cite[Section~6.4]{cheeyap}
\begin{equation}
  res(P, Q)=p_m^n q_n^m \prod_{i=1}^m\prod_{j=1}^n (a_i-b_j)
  \label{eq:resultant}
\end{equation}

\begin{figure*}
  \begin{align}
    \sign(a_1-b_1)=\sign(res(P, Q))\sign(p_m^n)\sign(q_n^m)
    \prod_{i=2}^m\prod_{j=2}^n[\sign(a_i-b_j)\sign(a_1-b_j)\sign(a_i-b_1)]
    \label{eq:signroot}
  \end{align}
\end{figure*}

The two expressions for the resultant provide us with another method
of computing the sign of one of the differences of the two roots.
Under our assumption that we know the order of all pairs or roots
except, say, $a_1$ and $b_1$, we can compute $\sign(a_1-b_1)$ from the
determinant and known signs, as in Equation \ref{eq:signroot} at the
top of the next page.  The signs need not be multiplied; we simply
count the negatives.  With quadratics, $m=n=2$, so the signs of the
leading ~$p_2^2$ and~$q_2^2$ will be positive and can be ignored.

The determinant can be computed with half the precision and fewer
floating point operations than repeated squaring to correctly compute
the sign of the differences of roots of the polynomials.

Computing a general $4{\times} 4$ determinant takes about 120
multiplications, and computing the determinant of the Sylvester matrix
itself would naively take~$35$ FLOPs for each comparison.  We can do
better in Equation~\ref{eq:sylvpoly} by writing the determinant in
terms of the discriminants and other precomputed $2{\times} 2$ minors
from each polynomial.  This brings us to 11 FLOPs per comparison, with
an initialization cost of 7 FLOPs per intersection.

\begin{figure*}
  \begin{equation*}
    \Delta=\begin{vmatrix}
    a_1 & b_1 & c_1 & 0\\
    0 & a_1 & b_1 & c_1\\
    a_2 & b_2 & c_2 & 0\\
    0 & a_2 & b_2 & c_2\\
    \end{vmatrix}=
    a_1^2 c_2^2 + c_1^2 a_2^2 + b_1^2 a_2 c_2 + b_2^2 a_1 c_1 -
    b_1 c_1 a_2 b_2 - a_1 b_1 b_2 c_2 - 2 a_1 c_1 a_2 c_2
  \end{equation*}
  \begin{align}
    \alpha_i=a_i^2,\,\, \gamma_i=c_i^2,\,\,
    \delta_i=a_i b_i,\,\, \epsilon_i=a_i c_i,\,\, \zeta_i=b_i c_i,\,\,
    D_i=b_i^2-\epsilon_i,\,\,
    i\in {1, 2}\\
    \Delta = \alpha_1 \gamma_2 + \gamma_1 \alpha_2 +
    D_1 \epsilon_2 + \epsilon_1 D_2 - \zeta_1 \delta_2 -
    \delta_1 \zeta_2
  \label{eq:sylvpoly}
  \end{align}
\end{figure*}

\section{Experimental Evaluation}
We experimentally evaluated the resultant method and the approximate
computation method with both machine precision and extended precision.
Repeated Squaring is dominated by the other methods so was not tested.

We created two types of test scenes that had touching surfaces so that
random lines might have some chance (albeit small) to give incorrect
orders under approximation, and count the number of disagreements.  We
evaluated time per comparison for each method on computers with
different processors.  Finally, by varying the number of surfaces in
the second type of scene, we could use linear regression to determine
the contribution to running time from per quadric and per comparison
terms.

\subsection{Experimental Setup}
All methods were implemented in C++, and were tested by computing the
line-quadric intersection orders along random lines in scenes of
quadric surfaces. The creation of these lines and quadric surfaces is
described in the next subsection.  Machine precision tests were
performed in IEEE 754, with quadratic coefficients and discriminants
stored as single precision floats, with all machine precision
computations performed as floats.  MPFR\cite{mpfr} was used to support
arbitrary precision in both the approximate comparison and the
resultant methods.  The approximate comparison method used $24\times$
the precision of a float.  The resultant comparison method also used
$24\times$ the precision of a float, to account for the range of
exponents in the inputs.

The first step of the evaluation for a line $\ell$ and quadric $Q$ was
to determined if there was a real intersection by evaluating the
discriminant of the quadratic~$p(t)=Q(\ell(t))$.  This evaluation was
done in machine precision, so there is a small chance that near
tangent intersections may have been missed due to numeric error in
calculating the discriminant. (In our application, missing near
tangent intersections was allowed, but getting orders wrong had been
known to trap particles into repeatedly trying to cross the same pair
of surfaces, which tends to worry a physicist.)

If the intersections are deemed to exist, the second step is to
compute the roots at machine precision.  These roots are needed to
determine if the order of a pair of intersections is ambiguous or not.
Finally, the stl sort algorithm is used to sort the intersections.
The full process was timed in nanoseconds with the POSIX
clock\_gettime function.

The comparison function used for sorting came from the method being
evaluated.  The machine precision approximate comparison just returns
the difference of the previously computed roots.  In the increased
precision approximation and the resultant method, the difference of
the roots is compared against a threshold.  If the difference was
smaller than a threshold of $2^{-16}$, the more accurate method
provided is used to determine the order, and an appropriate value is
returned.  This occurred infrequently for a random line, and is only
expected to occur a few times for every 100k lines.

We ran tests on two computers with different speeds and operating
systems; we name them by their operating systems.

\noindent{\bf Arch} was a Core i3 M370 processor with 2 cores, a 3 MB
cache, and 4 GB of DDR3 memory clocked at 1 GHz.  It ran an up-to-date
installation of Arch Linux, kernel version 4.4, and GCC 6.0 was used
to compile the code.  For the tests, the performance manager was set
to keep the CPU clock at 2.4 GHz, and the process was run with a nice
value of $-20$.

\noindent{\bf Gentoo} was a Core 2 Duo E6550 processor with two cores,
a 4 MB cache, and 8 GB of DDR2 memory clocked at 667 MHz.  It ran an
up-to-date installation of Gentoo Linux, kernel version 4.1 and GCC
4.9 was used to compile the code.  For the tests, the performance
manager was set to keep the CPU clock at 2.3 GHz, with a nice value of
$-20$.

A Geekbench benchmark was employed to estimate the floating point
processor speeds, Arch 1702, and Gentoo 1408. Thus, on average, Arch
was capable of about 1.2 times more FLOPS than the Gentoo computer.

\subsection{Test Scenes}
We created two types of test scenes: a single scene of Packed Spheres
and a set of scenes of Nested Spheres.  The test scenes consisted of
quadric surfaces stored as IEEE754 single precision floating point
numbers.  We preferred spheres and ellipsoids, since any intersecting
line would intersect twice, possibly with a repeated root.  Sorting
isolated single roots is easier, since, for example, the intersection
with a plane requires less precision. The quadric surfaces were
constructed from the unit cube that has one corner at the origin and
the opposite corner at $(1.0, 1.0, 1.0)$.

\begin{figure}
  \includegraphics[width=0.5\textwidth]{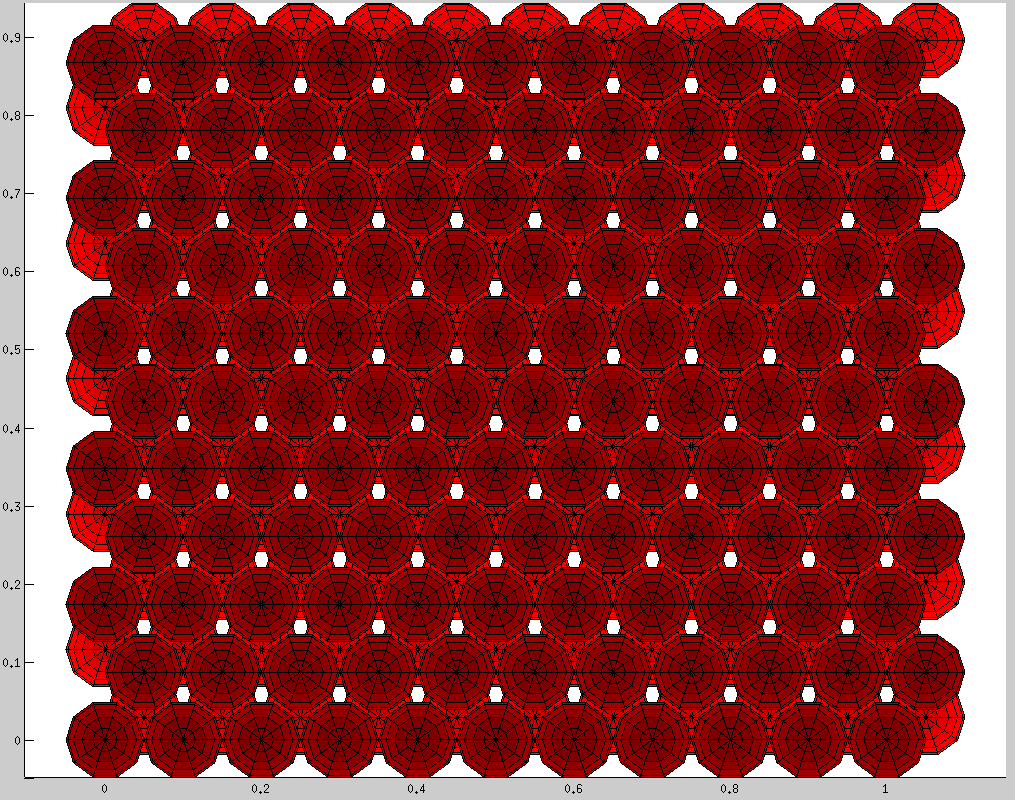}
  \vspace{5mm}
  
  \includegraphics[width=0.5\textwidth]{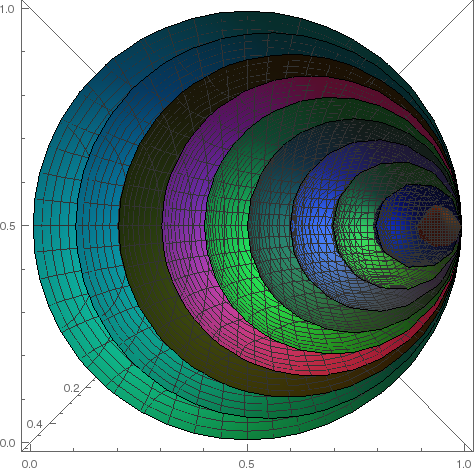}
  \caption{Test Scenes of 1331 Packed Spheres and 10 Nested Spheres,
    which is smallest of a family of eleven.  Random lines in Packed
    Spheres have some chance of being near sphere contacts.  Random
    lines in Nested Spheres are unlikely to, unless they are biased to
    pass by the near tangency.}
  \label{fig:testScenes}
\end{figure}

The single scene of Packed Spheres consisted of 1331 spheres in a
hexagonal close packing lattice shown in Fig.~\ref{fig:testScenes}.
This ensures that the spheres each have 12 intersecting or nearly
intersecting neighbors.  The spheres each have a radius of
about~$0.05$ units, and are spaced about~$0.05$ units from each other.
The initial sphere is centered at the origin, and one of the axes of
the lattice is aligned with the $y$ axis of the coordinate frame.  The
coefficients of the spheres are scaled so that the coefficients of the
squared terms were all $1.0$. This caused the exponent range for the
non-zero coefficients of the spheres to be between~$-8$ and~$1$, which
is well within the limits required for the resultant method to return
correct results.

The random lines generated for the scenes of Packed Spheres were
generated with an intersect from a uniform distribution over the unit
cube.  The directions were generated by normalizing a vector chosen
from a uniform distribution over the cube with opposite corners at
$(-1.0, -1.0, -1.0)$ and $(1.0, 1.0, 1.0)$.  To ensure that we are
able to compute the order of intersections exactly with the resultant
method, the exponents of the non-zero terms were constrained between
-20 and 0.

We used eleven scenes of Nested Spheres. One, shown in
Fig.~\ref{fig:testScenes}, had $n=10$ spheres, the others had
$n=100i$, for $1\leq i \leq 10$.  The first sphere was centered
at~$x_0=0.5, y_0=0.5, z_0=0.5$ units with a radius of~$R_0=0.5$ units.
The radius of successive spheres decreased linearly so that the final
sphere's radius was $R_n=2^{-16}$ units.  Thus,
$R_i=R_{i-1}-(R_0-R_n)/n$.  The $x$ position of successive spheres
increased linearly to fix the minimum distance at~$\epsilon=2^{-19}$
units. Thus, $x_i=x_{i-1}+(R_0-R_n)/n-\epsilon$.  The exponent range
for the non-zero coefficients of the spheres was chosen to be
between~$-1$ and~$0$, which is well within the limits required for the
resultant method to return correct results.

The random lines generated for the scenes of Nested Spheres were
generated with intersects~$p_i$ from a uniform distribution over the
unit cube.  The directions were set as~$(1.0, 0.5, 0.5)-p_i$, where
$(1.0, 0.5, 0.5)$ is a point very close to the points of minimum
distance for the sets of spheres.  This made it very probable that
increased precision would be required to correctly compute the order
of intersections.  To ensure that we are able to compute the order of
intersections exactly with the resultant method, the exponents of the
non-zero terms were constrained between -20 and 0.

\subsection{Analysis}
The time that it takes to compute the order of intersections between a
given line and a scene of quadric surfaces is expected to be linear in
both the number of quadric surfaces and the number of accurate
comparisons made.  Because performing accurate comparisons is so much
more expensive than normal comparisons, we expect there to be a clear
linear relation between the number of accurate comparisons performed
and the time it takes to perform the sorting.

The number of quadrics, on the other hand, can significantly affect
the number of intersections in the list to be sorted, especially in
antagonistic scenes.  However, most of the time spent sorting will be
accounted for by the time spent making accurate comparisons, which we
have already accounted for.  Thus, the remaining time will instead
come from computing the approximate roots, which is linear.

To analyze the Packed Spheres timing data, we used least squares to
fit a line to the number of comparisons made and the timing data.  A
constant term was also computed for the time taken computing the
approximate roots.

To analyze the set of Nested Spheres scenes, we aggregated the test
results for the scenes so that we could use least squares to fit a
plane to the number of comparisons made, the number of quadric
surfaces, and the timing data.  A constant term was also computed to
catch any hidden initialization costs, though we expect this to
contain mostly noise.

\section{Experimental Results}
The results of the experiments are shown in Table~\ref{tab:times}.
The first thing to notice is that increasing the precision of a
computation is not enough to guarantee that the result will be
computed correctly.  Despite increasing the precision of the
computations to~$24\times$ the initial precision, the increased
precision approximation still fails for~$1044/11000$ of the random
lines in the Nested Spheres scenes.  It did, however, perform
significantly better than the original calculation, which failed
for~$8272/11000$ of the lines.  More lines are needed to find examples
that cause errors in the Packed Spheres scene, but based on previous
experiments, we can expect several to occur by the
$100\text{k}^\text{th}$ test.

In addition to guaranteeing correctness, the resultant method also
performed well against the generic increased precision method.  For
the set of Nested Spheres scenes, it cost slightly more to compute the
order of intersections on a time per quadric basis.  The approximate
computation with increased precision can cache intermediate values
more effectively, reducing its cost.

The resultant method performed extremely well on the time per
comparison basis, as it actually beat the increased precision method
by more than it lost out on in the time per quadric basis in the
Nested Spheres scenes, and the Packed Spheres scene on the Gentoo
machine.

After removing the time per quadric basis in the tests with the Nested
Spheres scenes, the constant term appears somewhat nonsensical.  From
previous experiments, we have concluded that this is mostly noise,
suggesting that we obtained most of the useful information from the
measured times.  This suggests the time per quadric is the main
contributor to the constant time in the tests with the Packed Spheres
scene as we expected.

\begin{figure}
  \includegraphics[width=0.55\textwidth]{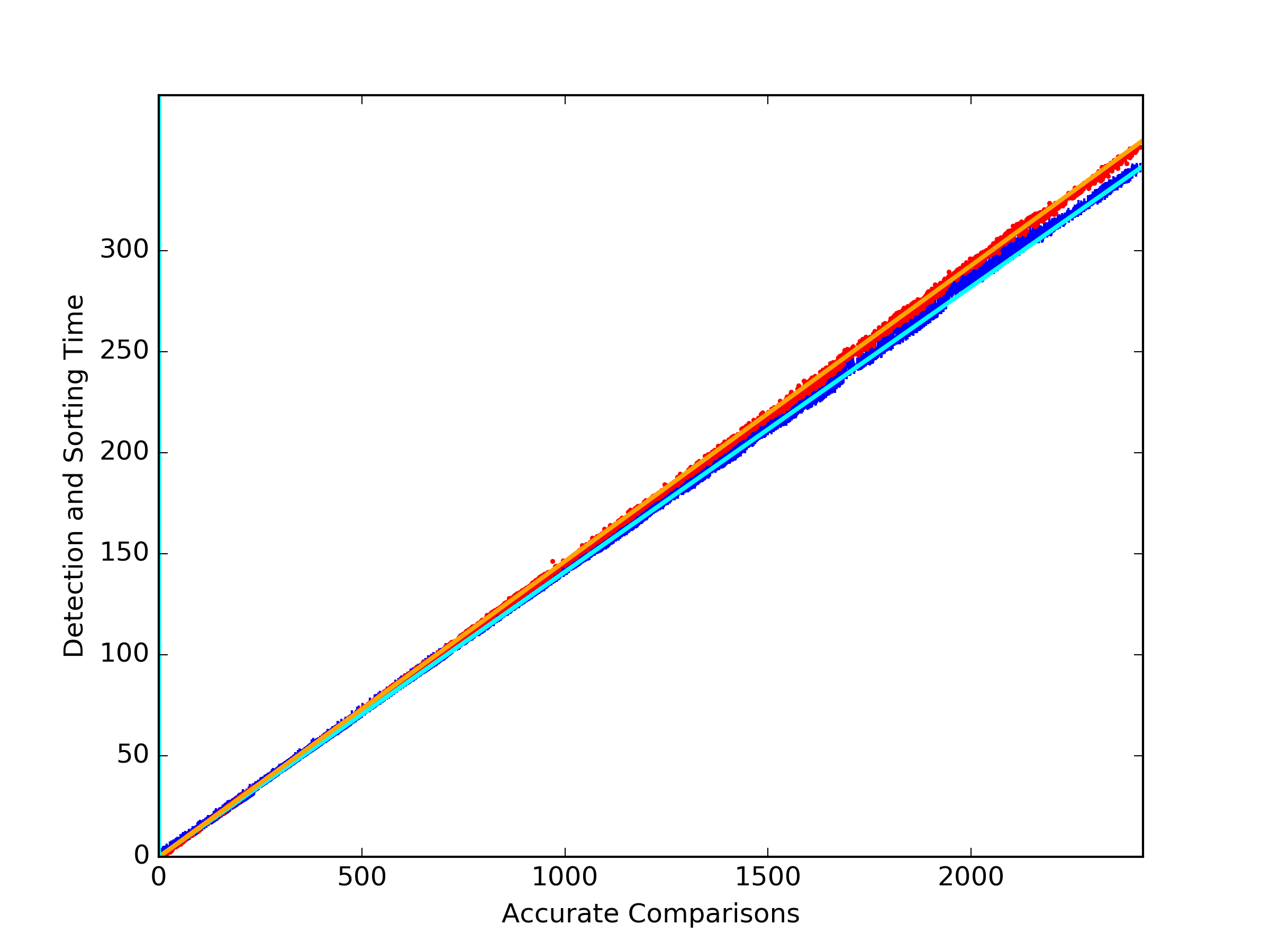}
  \caption{Evaluation Time for a Line (ms) vs. the Number of
    Comparisons; Sorting the intersections of 1k lines in each of the
    Nested Spheres test scenes on the Gentoo machine; {\bf Red Dot's
      (above the bars): Approximation Method at $24\times$ the Input
      Precision}; {\bf Blue Bars (beneath the dots): Resultant
      Method}.  The least squares coefficient for the time per
    quadrics has been subtracted out to better show the actual
    fit. The lines show the respective least squares fits without the
    quadric term.}
  \label{fig:linefit}
\end{figure}

Figure \ref{fig:linefit} shows a plot of the results from one of the
tests.  It appears to confirm our expectation that the time required
is linearly coorelated with the number of precision increases.

\begin{table*}
  \caption{Analysis of the timing of the Approximate Comparison and
    Resultant Comparison.  Timing data for 11k lines was analyzed for
    the Packed Spheres scene to find the coefficients of the best
    fitting lines.  Timing data for 1k lines was analyzed for each of
    the set of 11 Nested Spheres scenes to find the coefficients of
    the best fitting planes.  The dimensions are the number of quadric
    surfaces and the number of increased precision comparisons made.}
  \label{tab:times}
  \centering
  \begin{tabular}{|l|l|ll|lll|l|}
\hline
Scene & Machine & Method & Errors & ms/Quadric & ms/Comp & Const ms & $\sum$ Residual ($\text{ms}^2$)\\
\hhline{|=|=|==|===|=|}
Nested & Arch & Approximate & 8272 & 0.00425 & \hphantom{-}0.000361 & -0.0693 & \hphantom{000}\hphantom{-}149.084\\
Spheres &  & Increased Prec. & 1044 & 0.00554 & \hphantom{-}0.105 & -0.567 & \hphantom{0}\hphantom{-}87655.1\\
 &  & Resultant & \hphantom{---}--- & 0.00670 & \hphantom{-}0.100 & -0.746 & \hphantom{0}\hphantom{-}80544.6\\
\hhline{|~|-|--|---|-|}
 & Gentoo & Approximate & 8244 & 0.00379 & \hphantom{-}0.000313 & -0.0519 & \hphantom{0000}\hphantom{-}34.4705\\
 &  & Increased Prec. & 1042 & 0.00484 & \hphantom{-}0.146 & -0.110 & \hphantom{0}\hphantom{-}11944.9\\
 &  & Resultant & \hphantom{---}--- & 0.00584 & \hphantom{-}0.141 & \hphantom{-}0.00485 & \hphantom{0}\hphantom{-}19872.5\\
\hhline{|-|-|--|---|-|}
Packed & Arch & Approximate & \hphantom{000}0 & -- & \hphantom{-}0.00738 & \hphantom{-}4.54 & \hphantom{0000}\hphantom{-}21.7059\\
Spheres &  & Increased Prec. & \hphantom{000}0 & -- & \hphantom{-}0.126 & \hphantom{-}4.49 & \hphantom{0000}\hphantom{-}22.4387\\
 &  & Resultant & \hphantom{---}--- & -- & \hphantom{-}0.130 & \hphantom{-}4.51 & \hphantom{0000}\hphantom{-}23.5822\\
\hhline{|~|-|--|---|-|}
 & Gentoo & Approximate & \hphantom{000}0 & -- & \hphantom{-}0.00180 & \hphantom{-}4.37 & \hphantom{00000}\hphantom{-}3.75176\\
 &  & Increased Prec. & \hphantom{000}0 & -- & \hphantom{-}0.156 & \hphantom{-}4.37 & \hphantom{00000}\hphantom{-}3.76225\\
 &  & Resultant & \hphantom{---}--- & -- & \hphantom{-}0.155 & \hphantom{-}4.41 & \hphantom{00000}\hphantom{-}3.83604\\
\hline
\end{tabular}

\end{table*}

\section{Conclusion}
In this paper we showed how the resultant method can guarantee the
correct order of line-quadric intersections at a similar cost to using
an increased precision approximation method.  We have also shown that
naively using increased precision to improve accuracy is not enough to
eliminate errors, and that one must take into account the operations
being used and the ranges of the input.

We have assumed that we know the order of all roots except one
pair. Even if one's application does not provide this information, for
quadratic equations it is relatively easy to obtain using lower
precision than it takes to compare roots.  The zero of the derivative
$x_i=-b_i/a_i$ separates $r_i^-$ and $r_i^+$ by value of the
discriminant.  If $x_1=x_2$ then comparing squared discriminants tells
us all we need to know about root orders.  When, wlog, $x_1<x_2$, we
use the signs of both quadratics at $x_1$ and $x_2$ to bound roots to
intervals, and can again compare squared discriminants to reveal the
order for all but one pair.

\section{Acknowledgment}
We thank David Griesheimer for discussions on this problem, and both
NSF and Bettis Labs for their support of this research.

\bibliographystyle{plain}
\bibliography{resultantmethod}

\begin{thebibliography}{1}

\bibitem{brownawell2009lower}
W~Dale Brownawell and Chee~K Yap.
\newblock Lower bounds for zero-dimensional projections.
\newblock In {\em Proc. Int'l Symp on Symbolic and Algebraic Computation},
  pages 79--86. ACM, 2009.

\bibitem{emiris2010dmm}
Ioannis~Z Emiris, Bernard Mourrain, and Elias~P Tsigaridas.
\newblock The {DMM} bound: Multivariate (aggregate) separation bounds.
\newblock In {\em Proc. Int'l Symp on Symbolic and Algebraic Computation},
  pages 243--250. ACM, 2010.
\newblock \url{http://arxiv.org/abs/1005.5610}.

\bibitem{mpfr}
Laurent Fousse, Guillaume Hanrot, Vincent Lef\`{e}vre, Patrick P{\'e}lissier,
  and Paul Zimmermann.
\newblock Mpfr: A multiple-precision binary floating-point library with correct
  rounding.
\newblock {\em ACM Trans. Math. Softw.}, 33(2), June 2007.

\bibitem{cheeyap}
Chee-Keng Yap.
\newblock {\em Fundamental problems of algorithmic algebra}.
\newblock Oxford University Press, 2000.

\end{thebibliography}
\end{document}